\documentclass[twocolumn,showpacs,superscriptaddress,preprintnumbers,amsmath,amssymb,prc]{revtex4}

\usepackage{graphicx}
\usepackage{dcolumn}
\usepackage{bm}
\usepackage{indentfirst} 
\usepackage{color}
\usepackage{CJK}
\usepackage{booktabs}
\usepackage[colorlinks,linkcolor=blue,urlcolor=blue,citecolor=blue]{hyperref}

\begin{document}
\begin{CJK}{UTF8}{gbsn}

\title{Short range correlations  in the extended quantum molecular dynamics  model}

\author{Lei Shen(沈雷)}
\affiliation{School of Physical Science and Technology, ShanghaiTech University, Shanghai 201203, China}
\affiliation{Shanghai Institute of Applied Physics, Chinese Academy of Sciences, Shanghai 201800, China}
\affiliation{University of the Chinese Academy of Sciences, Beijing 100080, China}
\author{Bo-Song Huang(黄勃松)\footnote{Email: huangbosong@sinap.ac.cn}}

\affiliation{Shanghai Institute of Applied Physics, Chinese Academy of Sciences, Shanghai 201800, China}
\author{Yu-Gang Ma(马余刚)\footnote{Email: mayugang@fudan.edu.cn}}
\affiliation{Key Laboratory of Nuclear Physics and Ion-Beam Application (MOE), Institute of Modern Physics, Fudan University, Shanghai 200433, China}
\affiliation{Shanghai Research Center for Theoretical Nuclear Physics， NSFC and Fudan University, Shanghai 200438, China}

\date{\today}
\begin{abstract}
Short-range potential  has been added into an extended quantum molecular dynamics (EQMD) model. The RMS radius, binding energy and momentum distribution of $^{12}$C with different initial structures and short-range potential parameters have been checked. The separation energy of $^{12}$C(p,2p)$^{11}$B reaction is calculated and  compared with the experimental data, it indicates that our modified EQMD model can be taken as a reliable tool for studying proton pair knock-out reaction. Furthermore, the short range correlation effects on emission time and momentum spectrum of two protons are discussed. Finally, the momentum correlation function of the emitted proton pair is calculated by the Lednicky and Lyuboshitz's analytical method. The result explains that short-range repulsion causes high momentum tail and weakens the momentum correlation function.
\end{abstract}


\maketitle

\section{Introduction}

The short range correlation (SRC) is a very interesting topic recently in nuclear physics. SRC can partly arise from the nucleon-nucleon short-range central interaction. As a direct reflection of SRC, the momentum distribution of nucleons in nuclear system has been studied in many studies \cite{Czyz,Sartor,Antonov,Pandharipande,Egiyan,Frankfurt}. As a result caused by SRC, high momentum tail (HMT) can be found in the momentum distribution of nucleons. Some studies \cite{Hen,Ciofi,Hen2,Xu,Oertel}  show that the contribution of HMT to total wave function is about 20$\%$. In high energy electron scattering experiments at Jefferson Laboratory (JLAB) \cite{Hen,Subedi,Duer}  by measuring the relative abundances of neutron-proton pairs, neutron-neutron pairs and proton-proton pairs in nuclei from $^{12}$C to $^{208}$Pb, it is found that the neutron-proton pairs play a dominant role in SRC, and their ratio is much higher than the other two. 
Recently, a high-energy inverse kinematics ($p,2p$) scattering experiment has been performed to probe single-particle states and SRC in the well-understood $^{12}$C nucleus \cite{NatPhys}.
Some studies \cite{Cruz,West,LiBA,LiBA2}  show that there are spin and isospin dependences in nucleon-nucleon interaction, which are particularly important in SRC.

On the other hand, clustering behavior is an important nuclear structure phenomenon, especially for light nuclei, which can be observed in the excited state or even the ground state. Specifically, $\alpha$-cluster structure plays a leading role in even-number nuclei of $N = Z$. Ikeda proposed the threshold condition for the appearance of $\alpha$-clusters in $\alpha$-conjugated nuclei \cite{Ikeda}, which was also known as the famous Ikeda diagram. In recent years, the cluster structure in light nuclei has aroused great interests again due to the advances in nuclear theoretical methods and experimental techniques. 
The cluster structure near the threshold has a significant influence on the element abundance and nuclear synthesis in nuclear astrophysics, for example, the existence of the Hoyle state of $^{12}$C with $\alpha$ cluster structure has a decisive influence on the triple-$\alpha$ process and the abundance of $^{12}$C and $^{16}$O in constant stars \cite{Freer}. Because light nuclei are the main research objects of the first principles calculation methods, such as Green's function Monte Carlo method \cite{Wiringa}, core-shell model \cite{Navratil}  and effective field theory \cite{Epelbaum}, these methods try to reproduce the cluster properties in light nuclei by using realistic nucleon-nucleon interactions. 

In this work, we focus on SRC in $\alpha$-clustering nucleus, specifically for $^{12}$C. To this end, we added  the nucleon-nucleon repulsive potential in the framework of an extended quantum molecular dynamics (EQMD) model \cite{Maruyama}.  In this modified EQMD model, we studied the momentum spectrum, emission time distribution and momentum correlation function influenced by repulsive potential. Although short-range potential is not the whole cause of SRC effect, it is necessary to study the influence of short-range repulsive potential on the theoretical model.

The rest of this paper is arranged as follow: Section II provides a brief introduction of the original EQMD model and the newly-added repulsive potential as the SRC. Also the Lednicky and Lyuboshitz's analytical method for proton-proton momentum correlation is briefly described. In section III, we mainly discuss the calculation results of the emission time distribution, momentum spectrum and momentum correlation function of two emitted protons of $^{12}$C(p,2p)$^{11}$B reaction. Finally, a summary is given in Section IV.

\section{Model and Method}
\subsection{EQMD model}
Quantum molecular dynamics model (QMD) is an important tool to study the properties of atomic nucleus and simulate the process of nuclear collision \cite{Aichelin}. QMD model uses the direct product of Gaussian wave packet to express the wave function of the system. The canonical evolution equation of the system is obtained by variational method. After the introduction of coalescence method or coupling with statistical decay model, it has a good description of the nuclear fragmentation process \cite{He_NST,Guo,Yan1,Yan2,ZhangZF}. The traditional QMD model is more suitable for the description of medium heavy nuclei, and can give rich physical information for intermediate  energy nuclear reactions. However, it is not accurate to describe the properties of light nuclei, including those far from the stable line. In 1996, Maruyama ${\it et~ al.}$  \cite{Maruyama}  proposed an extended QMD model by introducing a series of new improvements, which is called EQMD model. In the  EQMD model, complex Gaussian wave packet and dynamic wave packet width were introduced to describe nucleons. The wave packet of the nucleon is described as
\begin{multline}
\phi_{i}(\vec{r}_{i}) = \bigg(\frac{v_i+v^{*}_{i}}{2\pi}\bigg)^{3/4}exp\bigg[-\frac{v_{i}}{2}(\vec{r}_{i}-\vec{R}_{i})^{2} +\frac{i}{\hbar}\vec{P}_{i}\cdot \vec{r}_{i}\bigg],
\label{Eqa_rhor}
\end{multline}
where $\vec{R}_{i}$ and $\vec{P}_{i}$ are the centers of position and momentum of the $i$-th wave packet, and the $v_{i}$ is the width of wave packets which can be presented as
${v_i} = {{1/{\lambda _i}}} + i{\delta _i}$  where $\lambda_i$ and $\delta_i$ are dynamical variables.
The ${v_i}$ of Gaussian wave packet for each nucleon is dynamical and independent.

\indent The wave packet expression in momentum space can be obtained by Fourier transform of single nucleon wave packet function in coordinate space
\begin{multline}
\phi_{i}(\vec{p}_{i}) = \bigg(\frac{v_i+v^{*}_{i}}{2\pi\hbar^{2}v^{2}_{i}}\bigg)^{3/4}exp\bigg[-\frac{1}{2v_i\hbar^{2}}(\vec{p}_{i}-\vec{P}_{i})^{2} -\frac{i}{\hbar}\vec{p}_{i}\cdot \vec{R}_{i}\bigg],
\end{multline}
and the density distribution function of a single nucleon can be obtained by making the complex conjugate inner product of the wave packet function.
The density distribution function in coordinate space can be described as
\begin{multline}
\rho(\vec{r}_{i}) = \phi_{i}(\vec{r}_{i})^{*}\phi_{i}(\vec{r}_{i}) = \bigg(\frac{1}{\pi\lambda_{i}}\bigg)^{3/2}exp\bigg[-\frac{1}{\lambda_{i}}(\vec{r}_{i}-\vec{R}_{i})^{2} \bigg],
\end{multline}
and while the  density distribution function in momentum space is expressed by 
\begin{multline}
\rho(\vec{p}_{i})=\phi_{i}(\vec{p}_{i})^{*}\phi_{i}(\vec{p}_{i})\\
=\bigg(\frac{1}{\pi}\cdot\frac{\lambda_{i}}{1+\lambda_{i}^{2}\delta_{i}^{2}}\bigg)^{3/2}
exp\bigg[-\frac{\lambda_{i}}{1+\lambda_{i}^{2}\delta_{i}^{2}}(\vec{p}_{i}-\vec{P}_{i})^{2} \bigg].
\label{Eqa_momentum}
\end{multline}

\indent In the EQMD, Pauli potential is introduced to make the ground state of the nucleus satisfy Fermion property, and the phase space state at the lowest energy  is obtained by a friction cooling method. 
Since in the common-used initialization method of QMD, the initial phase space information is obtained by the Monte Carlo method according to the uniform distribution in the space with radius $R$, so the nuclear phase space obtained by this direct random sampling is very unstable, and the initialization system is generally in excited states. In order to get a reasonable ground state or other stable states, a friction cooling process is needed in the EQMD. By introducing a damping term into the evolution equation, the system gradually cools down to a lower energy state with time evolution. Due to the above improvements, the EQMD model can well describe the properties of nuclear ground state, cluster structure and even halo structure of light nuclei, see eg. 
\cite{W.B.He,W.B.He2,Huang2021}.  Furthermore, the predicted $\alpha$-cluster structure has been successfully taken as an input for the studies of ultra-relativistic heavy-ion collisions \cite{ZhangS,LiYA,MaL,HeJJ,ChengYL,HeJJ2021}.  Some other developments, such as quasi-deuteron photonuclear reaction \cite{Huang12C,Huang16O,HuangBS-2019}  as well as hard photon production channel \cite{ShiCZ,ShiCZ2} have been also presented based on the EQMD model in our previous work.

\subsection{Short range potential}

The interaction potential in the EQMD model mainly consists of four terms, namely Skyrme potential, Coulomb potential, symmetry energy potential and Pauli potential, it   reads as 
\begin{equation}
H_{int} = H_{Skyrme} + H_{Coulomb} + H_{Symmetry} + H_{Pauli}.
\end{equation}
Here we mainly focus on the Skyrme potential, whose expression is
\begin{equation}
H_{Skyrme} = \frac{\alpha}{2\rho_0}\int\rho^2(\vec{r})d^{3}r+\frac{\beta}{(\gamma+1)\rho_0^\gamma}\int\rho^{\gamma+1}(\vec{r})d^{3}r. 
\end{equation}
\indent The first one is the two-body interaction potential, and the latter term is the three-body interaction potential with the parameter $\gamma$ = 2. According to the parameters given in Ref. \cite{Maruyama}, we  know that the two-body  interaction  potential in the Skyrme potential is attractive  and the three-body interaction  potential is repulsive.

\indent However, in the real case, the two-body interaction potential should contain a strong repulsive potential in the short range, so the HMT cannot be observed according to the current EQMD model.

\begin{figure}
\includegraphics[scale=0.3]{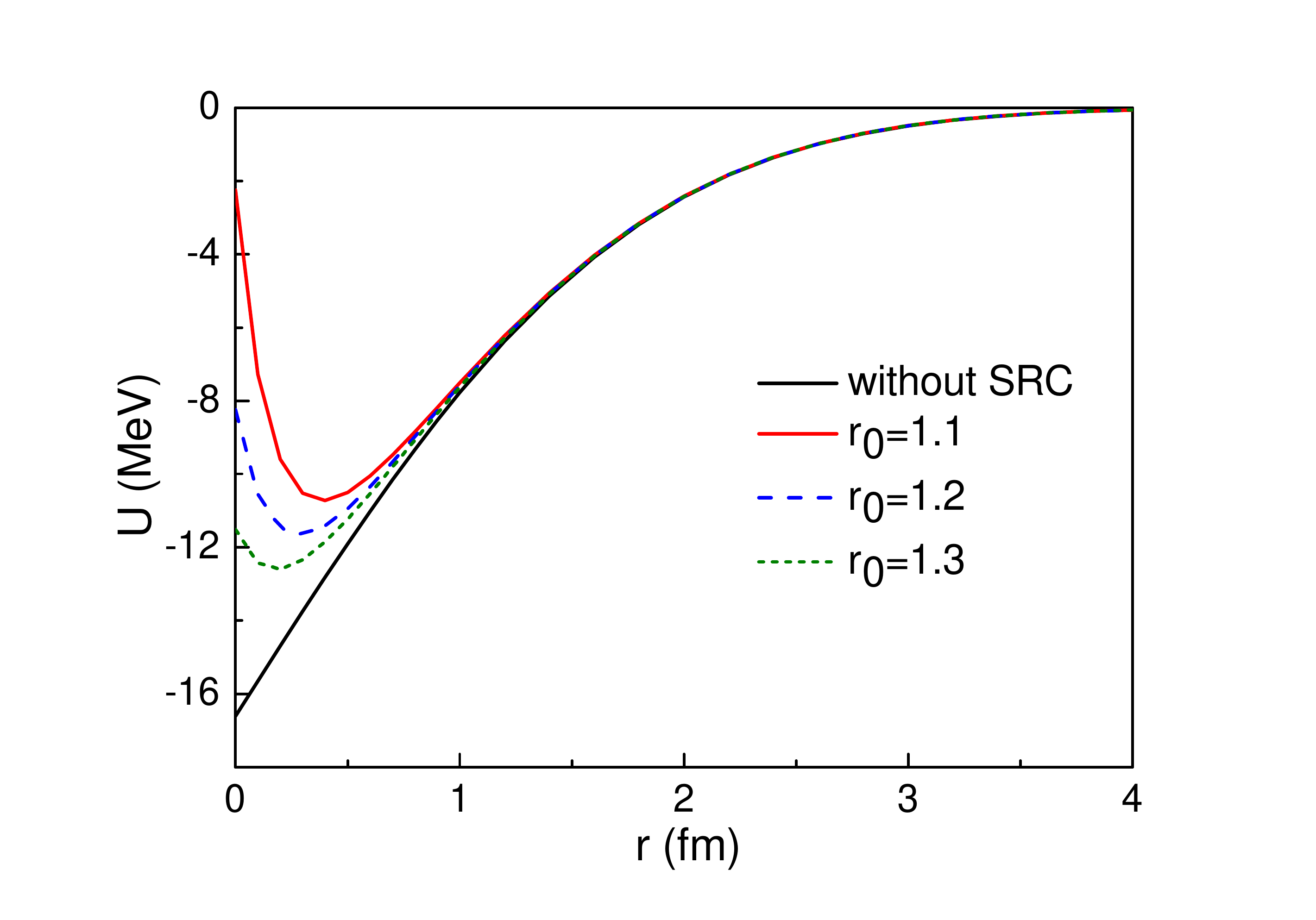}
\caption{The modified two-body potentials with different $r_0$, the black line is the original two-body potential $U_0$.}
\label{Fig_TBP}
\end{figure}

\indent Theoretically, when $r$ = 0, the repulsive potential should be infinite. However, it is not feasible in our model. So the method to consider the repulsive potential is to add the repulsive term which is similar to the so-called the Lennard-Jones potential \cite{Ranjan}  into the original two-body potential. The form is shown as follow
\begin{equation}
U = U_0 + U_1,
\end{equation}
\begin{equation}
\frac{dU_0}{dr} = -\frac{2C_{U_0}e^{\frac{-r_ij^2}{\lambda_i+\lambda_j}}}{\sqrt{\pi^3 (\lambda_i+\lambda_j)^5}},
\end{equation}
\begin{equation}
U_1 = \frac{C_{U_1}}{(r+r_0)^{p_1}},
\end{equation}
where $U_0$ is the original two-body potential of EQMD. The expression of $U_0$ is transcendental function, only the analytical expression of $\frac{dU_0}{dr}$ is given here. $r_{ij}$ is the distance of centers of wave packet of two nucleons. $U_1$ is the added repulsive potential. $C_{U_0}$ and $C_{U_1}$ are the derived constant coefficients of two potentials. In Fig. \ref{Fig_TBP}, we see that repulsive potential in short range is reduced with the increasing of the parameter $r_0$.

\indent Then we can calculate the influence of the adding SRC to the EQMD model. With different parameters the repulsive potential is finite at $r$ = 0 after setting $r_0$.

\subsection{Lednicky and Lyuboshitz's analytical method}

Some studies have shown that the proton pair momentum correlation function is an effective probe to study the SRC effect. Momentum correlation function, also known as the Hanbury-Brown and Twiss (HBT) effect \cite{HBT,Ma06,WangTT1,WangTT2,HuangBS-2019,Huang16O,Huang12C,HBT_LiLY,NST_Zhou}, is widely used in the study of heavy ion collision dynamics as well as the Fermion pair dynamics in three-body decay \cite{Lynch,Blank,Ma_PLB,Fang_PRC,WangSM1,WangSM2}. In this paper, Lednicky and Lyuboshitz's analytical method (LL model) \cite{Lednicky} is used to calculate momentum correlation function. The LL model can deal with the particle-particle correlation function with small relative momentum controlled by the quantum statistical effect and the final state interaction. The correlation function can be obtained by the sum of the squares of the mean Bethe-Salpeter amplitudes in the four coordinates of the emitted particles and the total spin of the two-particle system. Based on the conditions described in Ref.~\cite{Lednicky}, the correlation function of two particles can be written as
\begin{equation}
\vec{C}\left(\vec{k}^*\right) = \frac{\int \vec{S}\left(\vec{r}^*,\vec{k}^*\right)
\left|\Psi_{\vec{k}^*}\left(\vec{r}^*\right)\right|^{2}d^{4}\vec{r}^*}{\int\vec{S}\left(\vec{r}^*,\vec{k}^*\right)d^{4}\vec{r}^*},
\end{equation}
where $\vec{r}^* = \vec{x}_{1}-\vec{x}_{2}$ is the relative distance of two particles at their kinetic freeze-out, $\vec{k}^*$ is half of the relative momentum between two particles and later one we use $q$ for the same quantity, $\vec{S}\left(\vec{r}^*,\vec{k}^*\right)$ is the probability to emit a particle pair with given $\vec{r}^*$ and $\vec{k}^*$, $i.e.$, the source emission function, and $\Psi_{\vec{k}^*}\left(\vec{r}^*\right)$ is Bethe-Salpeter amplitude which can be approximated by the outer solution of the scattering problem \cite{Adamczyk}.

\section{Results and discussion}

After cooling process discussed above in the EQMD model, we obtain the stable target nucleus with short-range potential. With  different settings of the Pauli potential, we can obtain $^{12}$C nucleus with different structure, such as triangle $\alpha$-cluster structure and spherical distribution, namely the Woods-Saxon distribution. Using the equation (\ref{Eqa_rhor}), the RMS radius can be calculated. The results of $r_{RMS}$ and binding energy are shown in Table \ref{Tab_RMS} where the experimental data is taken from the IAEA nuclear data. Table \ref{Tab_RMS} indicates that the RMS radius of triangular cluster $^{12}$C is larger than the experimental data, while the one of spherical distribution $^{12}$C is lower than the data. The binding energy of triangular cluster $^{12}$C is lower than the experimental data, while the one of spherical distribution $^{12}$C is higher than the data.  We can consider that both the cluster structure and spherical distribution could be  in the ground state of the nucleus, but they are different components of the ground state. After superimposing the two results with a certain proportion, the result could be better close to the experimental data. Table \ref{Tab_RMS} also indicates that with the addition of stronger repulsive potential for the spherical configuration (i.e. from $r_0$ = 1.3 to $r_0$ = 1.1), the RMS radius becomes larger, but the binding energy becomes lower; for the triangular case, the RMS radius changes not much, but the binding energy becomes lower as well.\\
\begin{table}
\caption{\label{Tab_RMS}RMS radius and binding energy of $^{12}$C nucleus with different  structures. }
\begin{tabular}{|c|c|c|c|}
\hline
Without SRC & triangular & spherical & experimental\\
\hline 
$r_{RMS}$(fm)&2.57 & 2.30 & 2.47\\
$E_{bind}$(MeV)&7.26&8.71&7.68\\
\hline
Spherical with SRC& $r_0$=1.1 & $r_0$=1.2 & $r_0$=1.3\\
\hline
$r_{RMS}$(fm)& 2.37 & 2.34 & 2.32\\
$E_{bind}$(MeV)&6.88&7.59&8.00\\
\hline
Triangular with SRC& $r_0$=1.1 & $r_0$=1.2 & $r_0$=1.3\\
\hline
$r_{RMS}$(fm)& 2.35 & 2.37 & 2.34\\
$E_{bind}$(MeV)&6.12&6.61&6.87\\
\hline
\end{tabular}
\end{table}

\indent  
Momentum distribution of nucleons of $^{12}$C with SRC can be also obtained after cooling process as shown in Fig. 2 where the red,  blue and green lines indicate the momentum distribution with different parameters for repulsive potential. They show that the high momentum part of distribution increases with the addition of stronger repulsive potential, which is of course consistent with the SRC effect. 
 
\begin{figure}
\includegraphics[scale=0.3]{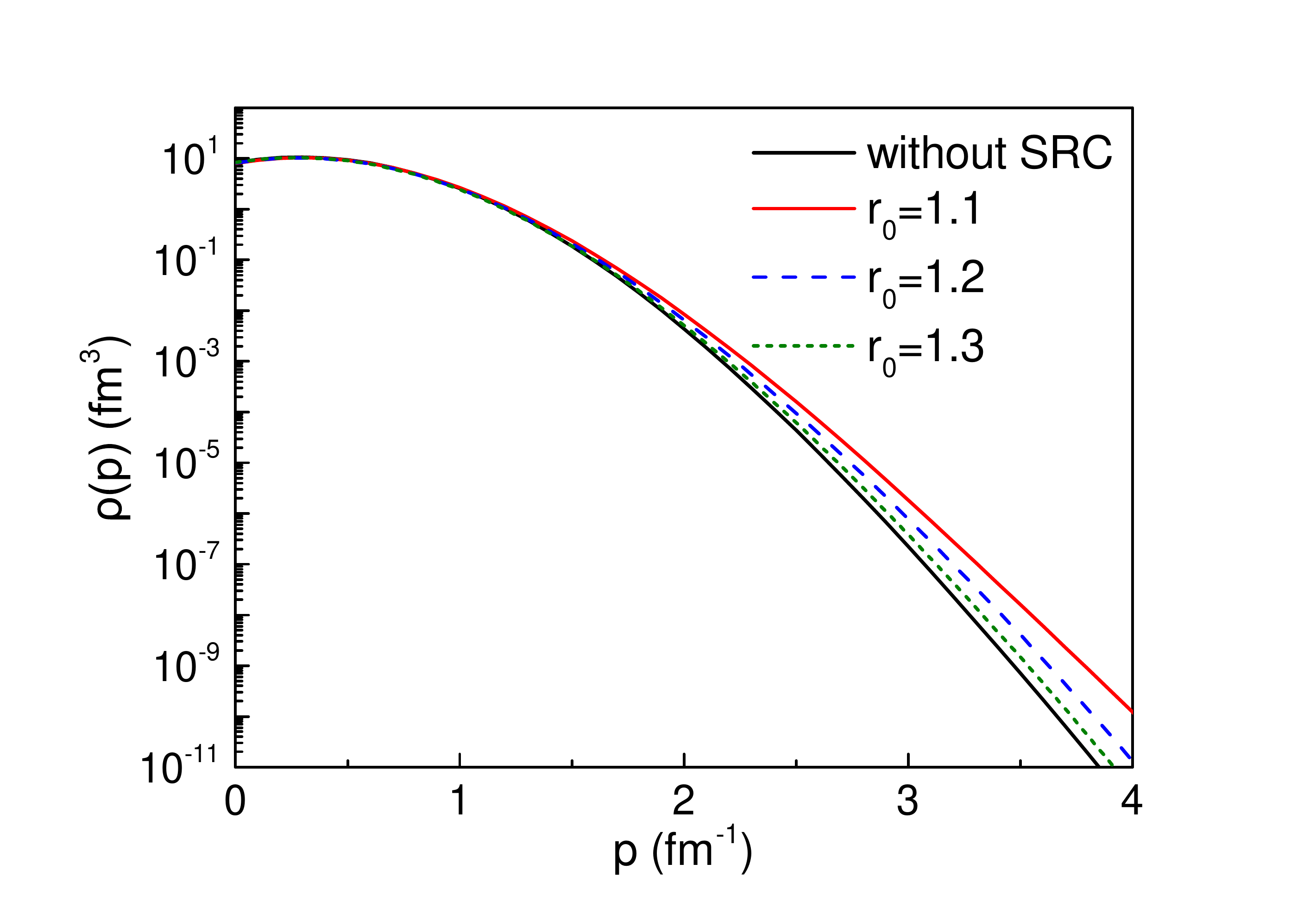}
\caption{Momentum distribution of nucleons in $^{12}$C after adding different repulsive potentials.}
\label{Fig_Momentum}
\end{figure}

\begin{figure}
\includegraphics[scale=0.3]{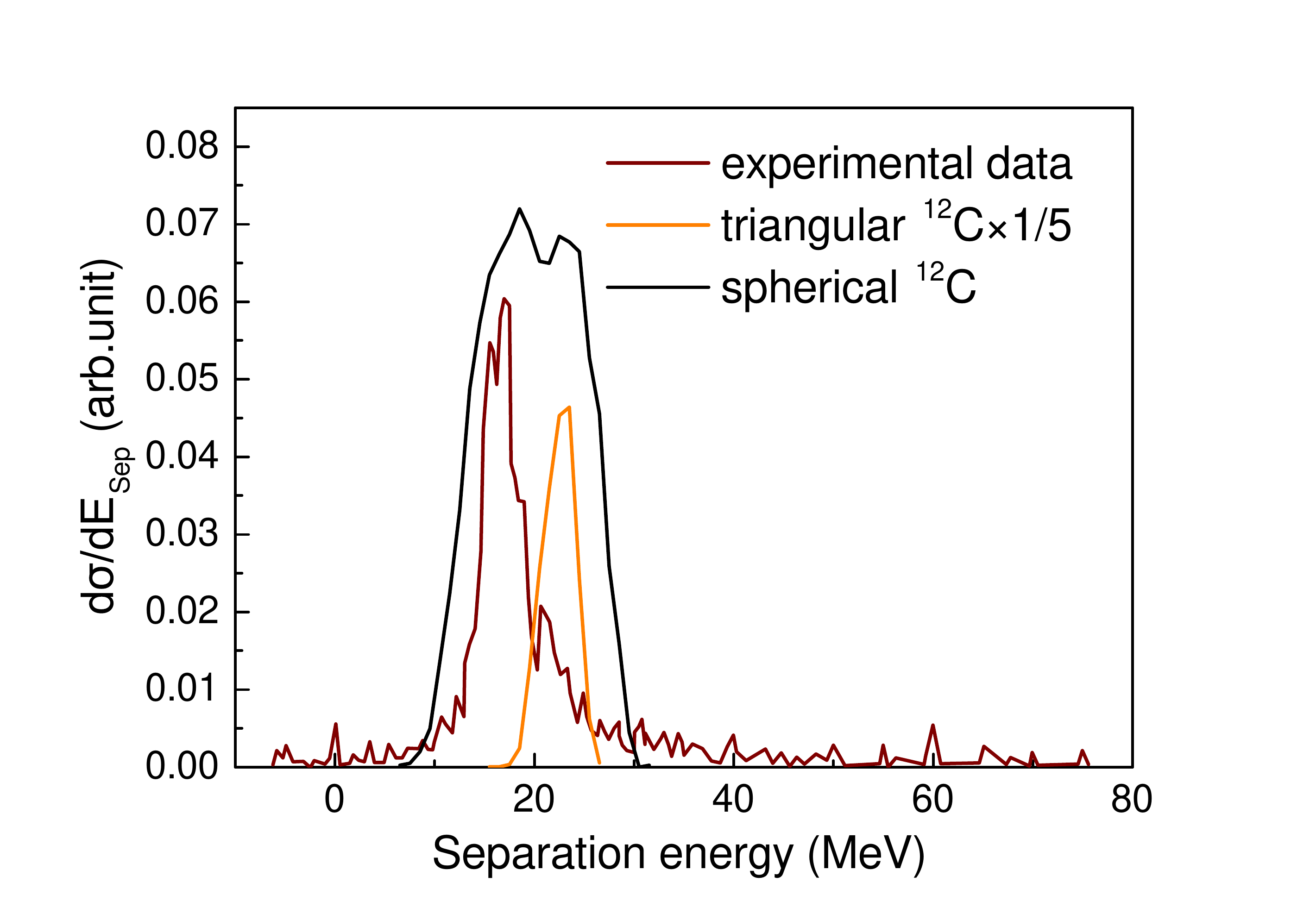}
\caption{Separation energy distribution of $^{12}$C(p,2p)$^{11}$B at 250 MeV. The experimental data is plotted by the dashed line, while  the spherical and triangle distribution simulations are shown by dark and orange lines, respectively.}
\label{Fig_sep}
\end{figure}

\indent It should be noted that the SRC effect may be also from other contributions such as tensor force \cite{Tensor1,Tensor2,Bai,Tensor3,Bai2}, which is not presented in the EQMD model. The tensor force mainly acts on spin-triplet, isospin-singlet neutron-proton pairs, and it significantly reduces the kinetic symmetry energy to even negative values at saturation density \cite{Tensor4,Tensor5,Tensor6,Tensor7}. The tensor force is spin and isospin dependent. The current EQMD model does not contain spin quantum numbers. Adding a new quantum number to EQMD model could be another study, which is not included in this work. Thus, the present modified EQMD model has limitation and thus room is open for further improvement in the future. In the present framework,  we  calculate the process of proton-target reaction. Here we only focus on the emission of two protons. Though the momentum distribution seems reasonable as mentioned above, we should also have  reliability check due to the addition of  repulsive potential in the present work. In order to demonstrate a reasonable result in our calculation intuitively, the process of $^{12}$C proton-pair knock-out at 250 MeV incident energy has been setup in our calculation and used for comparison with the experimental data. Here the target nucleus with two different initial structures, namely, triangular and spherical configurations, are respectively simulated. The separation energy of the process is calculated, which is compared with the experimental results of Kobayashi {\it et~ al.} in 2008 \cite{Kobayashi}, as shown in Fig. \ref{Fig_sep}.

\begin{figure}
     \includegraphics[scale=0.25]{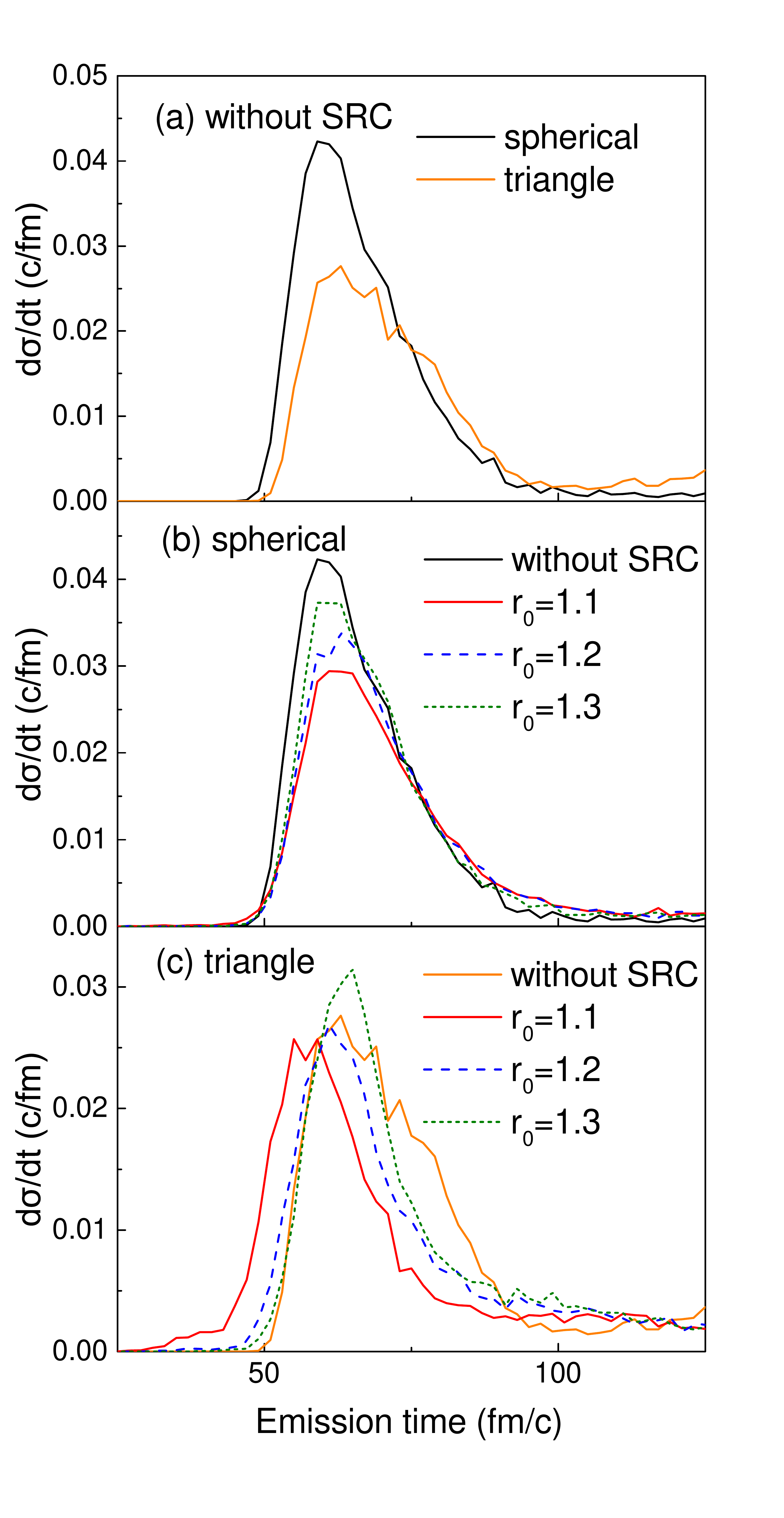}
    \caption{Emission time distribution of two ejected protons from $^{12}$C nucleus with different initial structure  in the case of w/o SRC (a);
     Comparisons of emission time distribution of the two emitted protons from the spherical (b) and triangular (c) $^{12}$C nucleus with different $r_0$ parameters.}
\label{Fig_emissiontime}
\end{figure}

\indent In the  figure, the orange line indicates the target nucleus $^{12}$C with three-$\alpha$ triangular structure while the dark one with spherical nucleon distribution. Here we use the area normalization, i.e. 1 for the spherical $^{12}$C and 1/5 for the triangular $^{12}$C. It shows that there is one peak in the separation energy spectrum  for triangular three-$\alpha$ structure, and  two peaks in the separation energy spectrum of $^{12}$C nucleus for the spherical nucleon distribution. The second peak is consistent with the peak of $^{12}$C nucleus with triangular structure. It is considered that both structures are not completely separated when screening the cluster structures of the initial nucleus after the cooling process, which leads to the possible existence of triangular structure in spherical nucleus. So we think that only the first peak is the separation energy spectrum of $^{12}$C nucleus for the Woods Saxon distribution. According to experiment by Bhowmik {\it et ~al.} in 1976 \cite{Ranjan}, different peaks will be formed in the separation energy spectrum according to the state of the nucleus, and the largest peak should correspond to the ground state of the nucleus. The EQMD model is a phenomenological transport model, which is not capable to simulate the excited states of nucleus very well. Therefore, the  highest point for the experimental data in the figure should correspond to the first peak of the separation energy spectrum of the Woods Saxon distribution. The peak value of the separation energy spectrum of $^{12}$C nucleus in the Woods Saxon distribution is slightly higher than the experimental data. Here we think that the main reason is that the EQMD model does not take into account the effect  of pion production. The incident energy of 250 MeV is higher than the threshold of pion production, and the pion generation will take away part of the energy. The separation energy calculation of the EQMD model without considering pion production effect  includes this part of energy, which leads to a slightly larger separation energy than the experimental value. The separation energy of $^{12}$C nucleus for the triangular three-$\alpha$ clustering structure is higher than that of $^{12}$C nucleus for the spherical nucleon distribution in Fig. 3, this is because that spherical distribution is more uniform, while the triangular cluster structure has three $\alpha$-clusters, and the $\alpha$ cluster binding energy is relatively higher than spherical distribution. The pair of protons knock-out reaction for triangular configuration is similar to the proton pair knock-out reaction from one $\alpha$ cluster, so the calculated separation energy of the triangular clustering $^{12}$C nucleus  is higher than that of the spherical distribution $^{12}$C nucleus. 
However, we notice that a small peak seems visible in the data at the same separation energy (around 20 MeV) as the triangle three-$\alpha$ configuration, which indicates that triangle three-$\alpha$ configuration is reasonable for explaining the separation energy in the region around 20 MeV.
In one word, the separation energy distribution of the $^{12}$C(p, 2p)$^{11}$B reaction simulated by the present modified EQMD model is approximately agreement with the experimental data, which indicates that the present model in which the short-range repulsive potential was taken into account is a suitable tool for studying the proton pair knockout reaction.

\begin{figure}
    \includegraphics[scale=0.25]{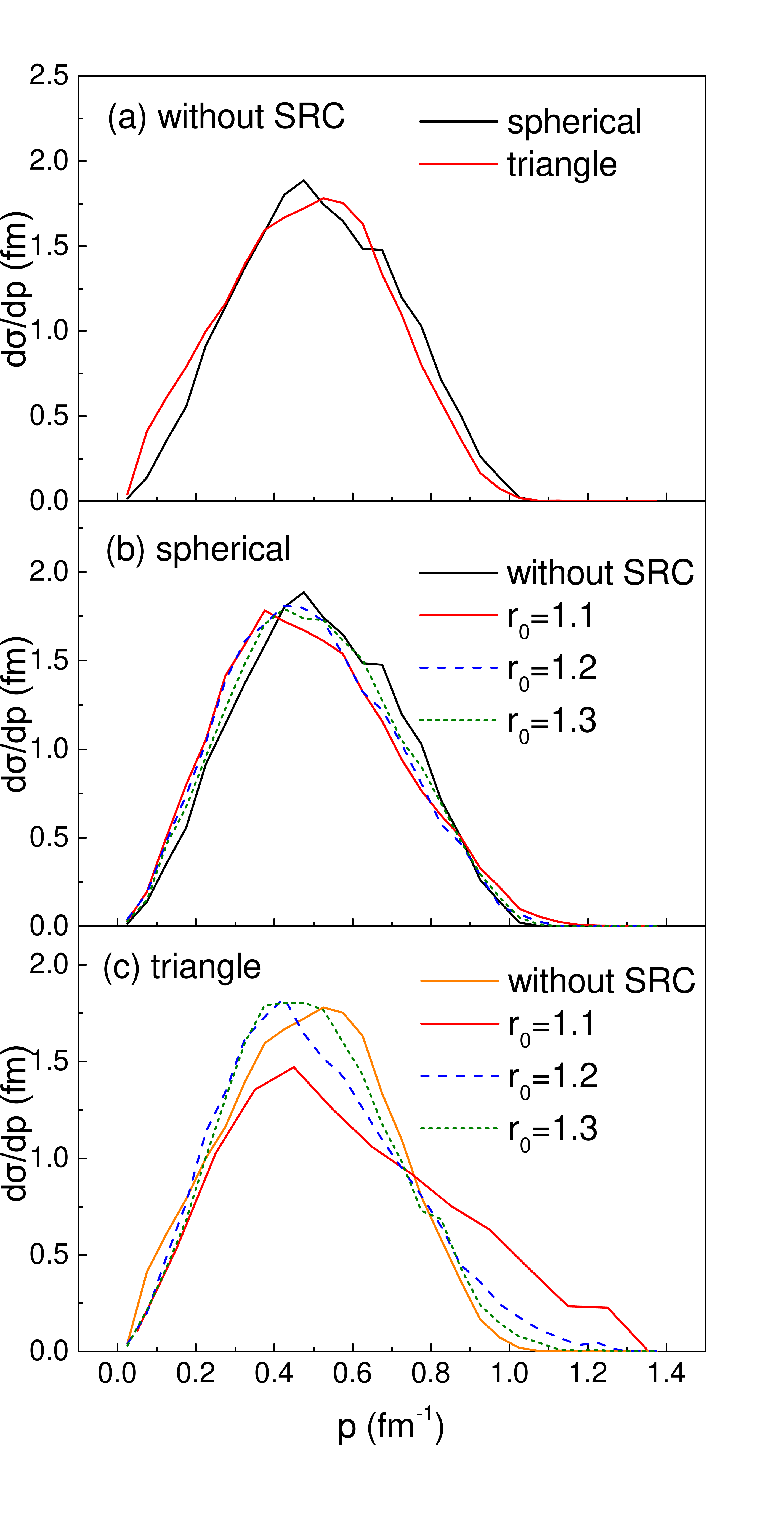}
    \caption{Same as Fig.~\ref{Fig_emissiontime} but for the momentum distribution of the two emitted protons.
    }
    \label{Fig_momentum2p}
\end{figure}

\begin{figure}
\includegraphics[scale=0.25]{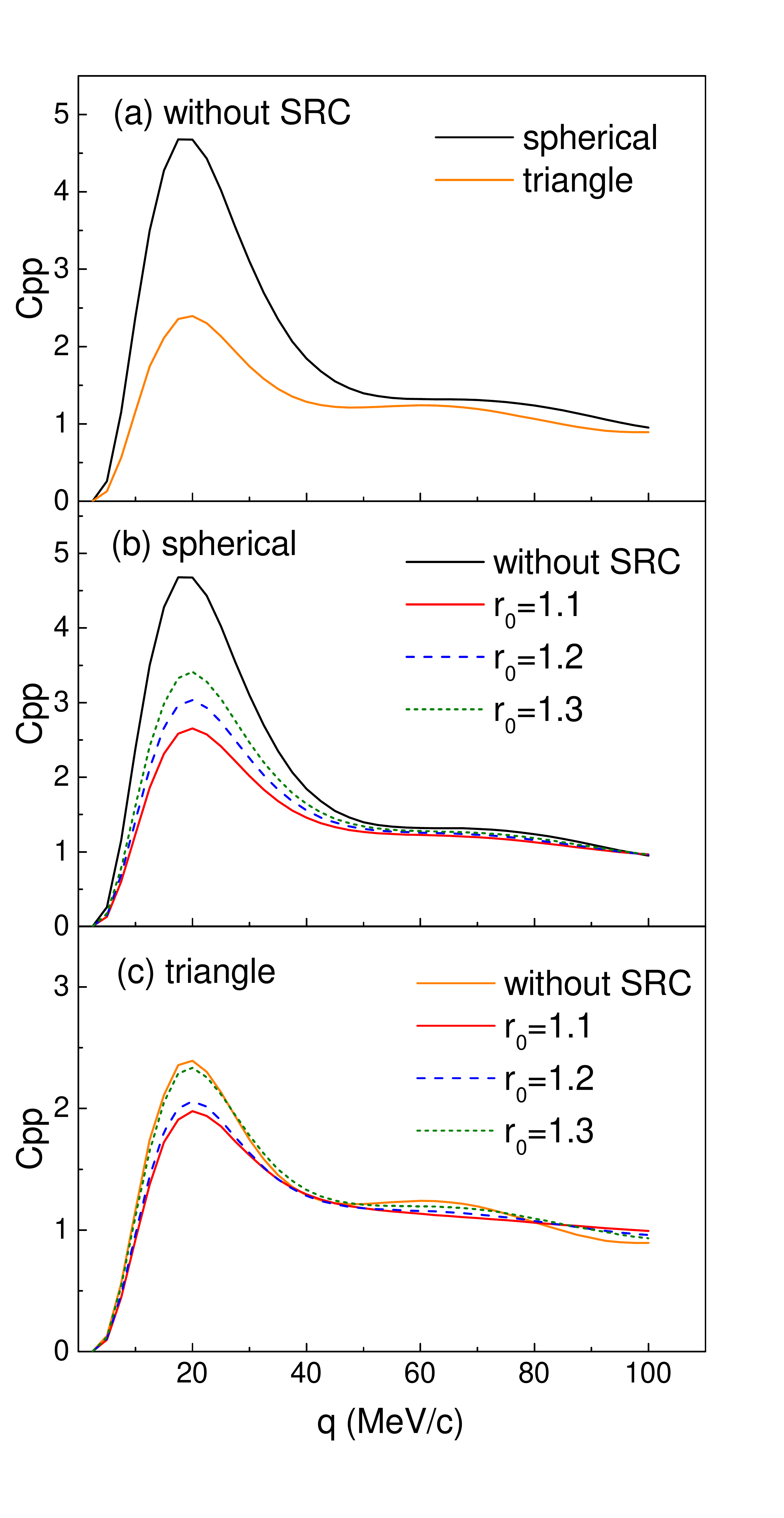}
\caption{Same as Fig.~\ref{Fig_emissiontime} but for the momentum correlation function of the two emitted protons.
}
\label{Fig_HBTcluster}
\end{figure}

\indent Based on the above check of separation energy spectrum, we can simulate the proton pair knock-out reaction to study the SRC effect. In this calculation, we only consider the channel of  two-proton emission, and  obtain the information of phase space and emission time when the two protons eject.

\indent Fig. \ref{Fig_emissiontime} 
shows the difference of emission time distribution of the two emitted protons in different situations. The results
show that most protons are emitted before 100 fm/c, which can be considered that they are caused by the knockout reaction. The other emitted protons are created in a uniform platform over 100 fm/c, which can be considered as a result of sequential decay. The higher the platform, the less stable the nucleus is. Thus, Fig. \ref{Fig_emissiontime} shows that the spherical distribution $^{12}$C nucleus is more stable than the triangular clustering $^{12}$C nucleus. Fig. \ref{Fig_emissiontime}(b) and (c) show that for both structures, the more SRC is added, the less stable the nucleus is. In addition, for triangle $^{12}$C case, the stronger SRC effect (red line, $r_0 = 1.1$) induces earlier emission of protons. 

\indent Fig. \ref{Fig_momentum2p} 
shows the difference of momentum distributions of the two emitted protons in different situations. Fig. \ref{Fig_momentum2p}(a) shows that the momentum distribution of the two emitted protons from $^{12}$C nucleus with different structures are almost the same. Fig. \ref{Fig_momentum2p}(b) and (c) show that the high momentum parts increase with the addition of stronger SRC, especially for the triangle case.

\indent Fig. \ref{Fig_emissiontime}(b) and (c) and Fig. \ref{Fig_momentum2p}(b) and (c) show that both of emission time distribution and momentum distribution of the triangular clustering $^{12}$C nucleus change more obviously with the addition of SRC. It is generally considered that the closer the nucleons are combined, the greater the influence of SRC. Table \ref{Tab_RMS} indicates that the RMS radius of the spherical distribution $^{12}$C nucleus is larger than the triangular clustering $^{12}$C nucleus, which means that globally the spherical distribution $^{12}$C nucleus combine closer than the triangular clustering $^{12}$C nucleus. We think that in each single $\alpha$ cluster the nucleons combine closer than the spherical structure, so the triangular clustering $^{12}$C nucleus is more sensitive to SRC.
For triangle $^{12}$C case, stronger SRC  potential (red line, $r_0 = 1.1$) induces obvious higher momentum  component. In other word, initial high momentum  tail can be somehow inherited by  the higher momentum component of ejected protons.
 
\indent Finally,  the momentum correlation function of the emitted proton pair can be calculated by taking this phase space and emission time information as the input of LL model which is described in section C. 
The calculation  result of  the proton pair knock-out reaction of $^{12}$C with different initial $^{12}$C configuration at 250 MeV is shown in Fig. \ref{Fig_HBTcluster}.
It shows that the momentum correlation as a function of relative momentum of the two emitted protons. There is a peak at $q$ = 20 MeV/$c$, which is due to the contribution of strong interaction as well as Coulomb interaction. The function then tends toward unity at larger relative momentum ($q$) because of the vanishing correlation. In Fig. \ref{Fig_HBTcluster}(a), we can see that the black line which refers to target nucleus with spherical nucleon distribution is significantly higher than the red line with the triangular cluster structure. The reason can be explained  by the effective emission source
size theory as Ref. \cite{HuangBS-2019}, which provides a similar result.

\indent The momentum correlation function of the proton knock-out reaction with the SRC effect is calculated by the same method, and the result is shown in Fig. \ref{Fig_HBTcluster}(b) and (c) for $^{12}$C nucleus with the spherical distribution and the triangular cluster structure, where the momentum correlation function calculated with different parameters $r_{0}$ for short range repulsive potential are displayed with green short dash line, blue dash line and red solid line, respectively, for $r_0$ = 1.3, 1.2 and 1.1, as well as the result  without the SRC effect which is depicted in black solid line.
 It is seen from Fig. \ref{Fig_HBTcluster}(a) that with the increase of the added repulsive potential, the peak gradually decreases, while the calculation without the short range repulsive potential gives the strongest correlation. Based on the explanation in Ref. \cite{HuangBS-2019}, the SRC leads to a larger size of the effective emission source, which leads to a lower peak. Besides, the increase of the natural decay part as shown in Fig. \ref{Fig_emissiontime}(b) and (c) can reduce the peak in the momentum correlation function, because there is no stable momentum or emission time correlation between randomly emitted natural decaying particles, which is also one of the explanations for the changing trend of the function. The result is also similar to the BUU result obtained in Ref.~\cite{Wei,Wei2}.

\section{Conclusion}

To summarize, we have embodied the short range repulsive potential into the EQMD model and checked the feasibility by binding energy and RMS radius. Then   the process of two-proton emission of proton-$^{12}$C reaction is focused  and the momentum correlation function of the emitted proton pairs are investigated.  Besides the initial target with a significant HMT is correctly depicted after the SRC repulsive potential is taken into account, we investigated the emission time distribution, the momentum spectra as well as the momentum correlation function with and without SRC. The calculation gives the trend that the larger the repulsive potential, the smaller the momentum correlation function. To some extents, the result explains that short-range repulsion causes HMT and leads to high momentum component of proton emission, but weakens the momentum correlation function of emitted protons.
Of course, the potential which was added in this model calculation is still preliminary. In order to obtain a more accurate result, we need to consider the tensor force. It is well known that the major origin of the tensor force is the one-pion exchange process which is not considered in this work. Besides, the spin and isospin wave function should be within this framework in further study. But anyway, the present work is still inspiring because of sensitivity of some observables to SRC as well as nuclear structure.

This work was supported in part  the National Natural Science Foundation of China under contract Nos.  11890710, 11890714, 11875066, 11925502, 11961141003 and 12147101, the Strategic Priority Research Program of CAS under Grant No. XDB34000000, National Key R\&D Program of China under Grant No. 2016YFE0100900 and 2018YFE0104600, and by Guangdong Major Project of Basic and Applied Basic Research No. 2020B0301030008.

\end{CJK}

\end{document}